# Sponsorship Disclosure in Native Advertising: A Theoretical Framework

Poompak Kusawat

Faculty of Commerce and Accountancy, Thammasat University, Bangkok, 10200, Thailand
Corresponding author E-Mail address: poompak-kus61@tbs.tu.ac.th

**Abstract**

Native advertising is one of the fastest growing areas of online promotion. After reviewing extant literature via EBSCOhost database, this study draws on Persuasion Knowledge Model and develops a theoretical framework which facilitates a clearer understanding of the relationship between sponsorship disclosure in native advertising and consumer outcome. The framework suggests that sponsorship disclosure has a negative effect on electronic word of mouth (eWOM), and further proposes the interplay between the main effect with brand prominence and the type of device. This is highly relevant to marketer as regulators have been pressuring for the disclosure of native advertising. As this is likely to have detrimental effect to the eWOM, marketer may employ the boundary conditions proposed by this framework to attenuate that negative effect.

**Introduction**

As long as its earliest days, media industry has been putting strong importance on the separation of content from advertising. Clear separation is seen as essential for maintaining the integrity of the content while isolating streams of revenue from advertising to sustain the business away from profit-driven pressure. However, in the last several years, the merging of two very separate but co-dependent entities have certainly evolved. This phenomenon is potentially caused by structural changes such as the increased advancement of digital media environment and abundance of information available to consumer in digital environment induces competition among advertisers for audience attention. Consequently, several alternative methods of advertising have emerged, including native advertising, which refers to any content that takes the appearance of the content from the publisher itself, the only difference is that it is paid for with the intent of advertising (Wojdynski & Evans, 2016). For example, Facebook ads content would appear in the news feed in a very identical form to organic post made by Facebook friends (Qiigo, 2019).

Native advertising is one of the fastest growing areas of online promotion. In 2021, native advertising is expected to drive 74% of all ad revenue (Boland, 2016). The growth of native advertising has sparked a debate about its effect on consumers (Sahni & Nair, 2016). On one side is the concern over consumers being misled to purchase a product they didn't mean to purchase due to organic nature of native advertising, resulting in advocates of regulations on native advertising. On the other side is the marketers who have effectively employed native advertising as well as the platform itself, since the regulations could potentially hamper with this prosperous emerging advertising market. Thus, deeper understanding about the nature of native advertising deserves researchers' attentions.

In December 2015, the Federal Trade Commission (FTC) issued guidelines and a policy statement relevant to native advertising. Specifically, it is regarding when disclosures in native advertising are required. For instance, Facebook ad displays its advertising nature by including the word "Sponsored" in the post, which supposedly cues the consumers that this post is paid to be shown on their news feed for advertising purpose. generally, people preferred not to be manipulated and want to independently make their decisions (Brehm &

Brehm, 1981). Therefore, it is presumed that people are likely to resist persuasion attempts when they recognize them (Petty & Cacioppo, 1977; Wei, Fischer, & Main, 2008), as such, disclosing the nature of native advertising then likely leads to negative outcome. Thus, it is practically necessary to find a way to counter this effect.

The objective of this paper is to provide a theoretical framework illustrating the effect of sponsorship disclosure in native advertising on consumer outcome, the mechanism underlying the effect, as well as the boundary conditions of the effect. Particularly, the proposed theoretical framework attempts to answer the following research questions:

1) What is the effect of sponsorship disclosure on electronic word of mouth (eWOM)?
2) Is persuasion knowledge the mechanism underlying the relationship?
3) What are the moderating effects of brand prominence and the type of device on the main relationship?

This paper adds to the existing literature in three ways. First, despite the arrays of studies on sponsorship disclosure, studies incorporating brand prominence is remarkably scarce. In addition, though studies on brand prominence can commonly be found in the extant literature, findings in general context may not hold for native advertising (Harms, Bijmolt, & Hoekstra, 2017). This paper provides a theoretical framework facilitating research on the boundary conditions of sponsorship disclosure by examining its interaction with brand prominence.

Second, in addition to brand prominence, sponsorship disclosure studies incorporating differences between device is also rarely found. Thus, the proposed framework aims to delve deeper and explore the interaction effect between sponsorship disclosure and the type of device used.

Third, this study adds to the literature on eWOM by examining a mechanism in which consumers' potential to share the advertisement is triggered. It is important for business to understand the mechanism that lead consumers to engage in eWOM because in the current era of digitization, marketers rely on eWOM as a key measure for advertising performance (Peters, Chen, Kaplan, Ognibeni, & Pauwels, 2013). Furthermore, it is generally found that eWOM positively influence firm performance (Liu, 2006; Sonnier, McAlister, & Rutz, 2011; Tirunillai & Tellis, 2012; Li & Hitt, 2008; You, Vadakkepatt, & Joshi, 2015; Yang, Kim, Amblee, & Jeong, 2012).

The remaining of this paper first discusses the relevant literature, then presents the propositions. The last section concludes.

## Literature Review

The literature search on native advertising was performed within EBSCOhost database, which is a fee-based online research service providing full-text databases.

### Native advertising

The cost and benefit associated with native advertising is continuing to be the subject of endless debate. Due to its similarity with the page content and its assimilation with the platform design, native advertising has significantly contributed to advertising success. However, concerns over native advertising's potential to misguide or deceive consumers have also emerged (Kim, 2017). Nevertheless, there exist various literatures attempting to study the effectiveness of native advertising.

The benefit of native advertising is likely due to its subtle nature. As one study found that after the Privacy Directive in the European Union (EU) was passed, advertising effectiveness decreased on average by around 65% (Goldfarb & Tucker, 2011), one possible explanation is that sponsorship disclosure required by this law may frame native advertising as a form of deception, which leads to negative evaluations (Darke & Ritchie, 2007).

**Consequences of native advertising**

Academics have studied numerous contexts in which the impact of sponsorship disclosure on advertising effectiveness may vary. For instance, extant literature has examined the effects of duration of disclosure (Boerman, van Reijmersdal, & Neijens, 2012), timing of disclosure (Boerman, van Reijmersdal, & Neijens, 2014), the type of deceived attribute (Held & Germelmann, 2014), the type of disclosure (Boerman, van Reijmersdal, & Neijens, 2015; Sahni & Nair, 2016), position of disclosure (Wang, Xiong, & Yang, 2019; Wojdynski & Evans, 2016), disclosure source (Boerman, Willemsen, & Van Der Aa, 2017), and brand presence (Krouwer, Poels, & Paulussen, 2017).

**Persuasion knowledge as the mechanism**

On the other hand, some studies focus on the mechanisms in which sponsorship disclosure affects consumers' response, the proposed mechanisms are visual attention (Boerman et al., 2015; Wojdynski & Evans, 2016), advertising recognition (Wojdynski & Evans, 2016), and persuasion knowledge (Boerman et al., 2012; Boerman et al., 2014; Boerman et al., 2017; Krouwer et al., 2017).

More closely related to this research are studies that consider the relationship of sponsorship disclosure and persuasion knowledge. Boerman et al. (2012) studied in television context, the impact of duration of sponsorship disclosure on brand response. They found that 6-second disclosure, as opposed to 3-second disclosure or no disclosure, activates persuasion knowledge, and ultimately affects brand attitude and brand memory. Boerman et al. (2017) took another perspective and examined the impact of the source of sponsorship disclosure on electronic word of mouth. They found that the effect of sponsorship disclosure on persuasion knowledge is salient only when the post is generated by a celebrity, but not by the firm itself. The reason is that the advertising nature of posts by firm is already within consumers' expectation, however, when the post is made by a celebrity, consumers expect a more organic content resulting in effective sponsorship disclosure.

Despite the extensive literature, no conclusive factors or mechanisms were yet suggested to both able to signal to consumers the advertising nature of the native advertising as required by laws and ethics as well as maintaining positive consumers' response at the same time. This suggests that alternative means of achieving disclosure is warranted.

## Research Propositions

**Electronic word of mouth**

eWOM refers to a positive or negative statement about a product, service, brand, or company. And that statement is produced by consumers and made publicly available through web-based technologies such as social media, websites, or internet forums (Hennig-Thurau, Gwinner, Walsh, & Gremler, 2004). Generally, consumers are willing to disseminate information through eWOM only when they believe that the information is accurate. Because they prefer to present themselves in a positive way so they avoid disseminating inaccurate information (Berger, 2014). Advocating on false facts about a product in eWOM would be perceived in a negative view. As a result, consumers are more likely to promote a product or brand only when they trust its authenticity. de Matos & Rossi (2008) supports this argument through a meta-analysis on numerous antecedents of face-to-face word of mouth. Chu & Kim (2011) validated these effects and extending it in the context of electronic equivalent of word of mouth. They found that trust is positively related to intention to engage in eWOM.

In the context of Facebook, sponsored Facebook posts are integrated into a user's Facebook newsfeed in an organic manner. That is, sponsored Facebook posts resemble those of everyday users'. Though this appearance is susceptible to confusion as consumers may or may not recognize the post as advertising, sponsored posts actually include a signal that discloses to consumers that the posts are made with commercial intent. When consumers notice a post with sponsorship disclosure and recognize the post as advertising, they negatively view the advertising as untrustworthy and their intention to share is diminished. In line with this argument, prior research has found that if consumers recognize the persuasive intent of the advertising, they tend to be less willing to share an advertising campaign on online videos (Hsieh, Hsieh, & Tang, 2012) and a social networking site (van Noort, Antheunis, & van Reijmersdal, 2012). Thus, I propose that:

$P_1$: **Sponsorship disclosure has a negative effect on eWOM.**

**Persuasion knowledge**

Persuasion knowledge is the knowledge and beliefs about issues concerning advertising. For example, the goals the marketers are pursuing, the marketing tactics used to persuade the consumers, consumers' perception about appropriateness of these tactics, as well as consumers' numerous ways to cope with these goals and tactics (Friestad & Wright, 1994; Hibbert, Smith, Davies, & Ireland, 2007). This knowledge consists of two dimensions: a cognitive and an affective dimension (Boerman et al., 2012; Rozendaal, Lapierre, van Reijmersdal, & Buijzen, 2011). The cognitive dimension of persuasion knowledge, also known as conceptual persuasion knowledge, is the first step of persuasion knowledge. It is consumers' understanding of the intent of marketers along with the tactics used for persuasion. Attitudinal persuasion knowledge is the affective dimension of persuasion knowledge. It considers consumers' tendency to disbelief or dislike advertising (Boerman et al., 2012; Rozendaal et al., 2011). The Persuasion Knowledge Model suggests that when consumers recognize a message's persuasion purpose, the beliefs about the appropriateness and fairness of this message and the tactics used are developed (Friestad & Wright, 1994).

Since sponsored Facebook posts looks almost indistinguishable from organic posts, in contrast to traditional advertising, sponsored Facebook posts should lead to lower level of persuasion knowledge. However, with the existence of sponsorship disclosure, consumers who observe the word 'Sponsored' in the sponsored post could associate the post as advertising and their conceptual persuasion knowledge is activated. And, when consumers recognize that a message contains a persuasive intent, they develop distrust towards the message and their attitudinal persuasion knowledge is increased (Boerman et al., 2012; Wei et al., 2008; Wood & Quinn, 2003). Thus, based on aforementioned reasoning and findings in the rich literature:

$P_2$: **The effect of sponsorship disclosure on eWOM is sequentially mediated by conceptual persuasion knowledge and attitudinal persuasion knowledge.**

**Brand prominence**

Brand prominence is defined as the way brands are integrated in the advertising content (Homer, 2009). The brands can be placed in the advertising either in a subtle manner or a prominent manner. In a subtle manner, the brands would appear small and placed closer to the back of the scene. However, in a prominent manner, the brands would be bigger and more central to the scene's foreground (Gupta & Lord, 1998; Kozary & Baxter, 2010). When brand is placed prominently, it becomes more noticeable by consumers and might be recognized as commercial content. Thus, brand prominence could lead to the activation of conceptual persuasion knowledge (Wei et al., 2008). However, even if both sponsorship disclosure and brand prominence activate consumers' conceptual persuasion knowledge, the effect may not be as straightforward as the whole is equal to the sum of its parts.

Whether the brand is prominently displayed or subtly displayed in a post, consumers' inference about advertising intent of the post is merely speculations by consumers. They cannot be certain that this post is an advertising. However, in the case of sponsorship disclosure, if consumers are aware of such revelation, they can assume for certain that this post is of advertising nature.

Based on previous line of argument, when there already is a sponsorship disclosure and given that consumers are aware of it, brand prominence should have no further effect on advertising recognition, thus whether the brand is being prominently placed or subtly placed should be of no difference in regards to consumers' conceptual persuasion knowledge. On the other hand, when sponsorship disclosure is not present, brand prominence should serve as a cue for advertising nature of the post. To summarize, posts with prominently placed brand are more likely to activate consumers' conceptual persuasion knowledge compare to posts with subtly placed brand only when there is no sponsorship disclosure.

**$P_3$: The effect of brand prominence on the activation of consumers' conceptual persuasion knowledge is more salient when there's no sponsorship disclosure.**

**Device type**

Though sponsorship disclosure should result in higher level of consumers' conceptual persuasion knowledge, the effect may vary in different context. The degree to which consumers pay attention to sponsorship disclosure most likely affects the likelihood that they will recognize it as advertising. Boerman et al. (2015) has shown that a disclosure positively affects the recognition of advertising through viewers' attention to the disclosure. Specifically, the effect of disclosure on the recognition of advertising is salient only when viewers pay attention to it.

Consumers' attention to sponsorship disclosure should in turn be influenced by the type of device used. Harms et al. (2017) suggested that a mobile device, as opposed to a desktop, provides minimal distraction relevant to other advertising content due to its smaller screens. Thus, consumers should be more likely to pay greater attention on sponsorship disclosure on a mobile device relative to a desktop. Consequently, this greater attention results in consumers' higher chance to recognize the content as advertising. Therefore, I propose that:

**$P_4$: The effect of brand sponsorship disclosure on the activation of consumers' conceptual persuasion knowledge is more salient on a mobile device relative to on a desktop.**

Figure 1 depicts the model for understanding the relationship between sponsorship disclosure in native advertising and consumer outcome. It is worth noting that this framework is applicable at individual consumer level.

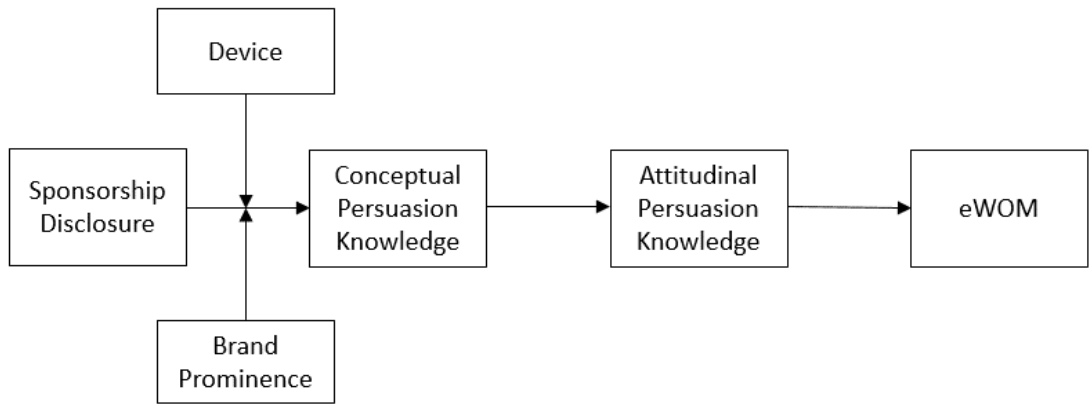

**Figure 1** Conceptual Framework

**Conclusion and Discussion**

This paper has set forward a number of research propositions along with the proposed theoretical model. Clearly, the propositions are in need of empirical investigation.

The proposed research agenda has three important theoretical contributions. First and foremost, this framework will shed light into the boundary conditions of sponsorship disclosure by examining its interaction with brand prominence. Although the concept of brand prominence has been around in marketing research for some time, it has not been given sufficient attention in the native advertising literature. I hope to fill this gap by examining the interaction effect of sponsorship and brand prominence. I propose that when there already is a sponsorship disclosure, brand prominence should have no further effect on advertising recognition. On the other hand, when sponsorship disclosure is not present, brand prominence should serve as a cue for advertising nature of the post. Secondly, similar to brand prominence, sponsorship disclosure studies incorporating differences between device is also hardly found. I hope to fill this gap by proposing that on a mobile device, consumers are more likely to pay greater attention to sponsorship disclosure due to less disturbance in smaller screen, resulting in consumers' higher chance to recognize the content as advertising compare to on a desktop. Third, this study adds to the literature on eWOM by examining a mechanism in which consumers' potential to share the advertisement is triggered. It is important for business to understand the mechanism that lead consumers to engage in eWOM because eWOM is generally found to positively influence firm performance, and reliance on eWOM as a key measure for advertising performance is becoming prevalent. Integrating the effect of brand prominence and device type with the current sponsorship disclosure literature would further enrich our understanding.

This framework is highly relevant to marketer. As there has been growing trends in native advertising, regulators have been pressuring for the disclosure of native advertising. However, as this is likely to have detrimental effect to the eWOM, marketer may employ the boundary conditions proposed by this framework to attenuate that negative effect. For instance, since the effect of sponsorship disclosure is more salient on mobile device, marketer may focus more resources on advertising in the desktop computer platform to optimize the benefit.